\documentclass{aastex}
\usepackage[onecolumn]{emulateapj5}

\newcommand{\ie}{i.e.,}
\newcommand{\eg}{e.g.,}

\newcommand{\etal}{et~al.}
\newcommand{\ltsima}{$\; \buildrel < \over \sim \;$}
\newcommand{\simlt}{\lower.5ex\hbox{\ltsima}}
\newcommand{\gtsima}{$\; \buildrel > \over \sim \;$}
\newcommand{\simgt}{\lower.5ex\hbox{\gtsima}}
\newcommand{\kms}{km~s$^{-1}$}

\begin{document}

\title{Kinematics of Planetary Nebulae in M51's Tidal
Debris}

\author{Patrick R. Durrell}
\email{pdurrell@astro.psu.edu} \affil{Department of Astronomy and
Astrophysics,
Penn State University, 525 Davey Lab, University Park, PA 16802}

\author{J. Christopher Mihos\altaffilmark{1}, John J. Feldmeier}
\email{hos@burro.astr.cwru.edu, johnf@eor.astr.cwru.edu}
\affil{Department of Astronomy, Case Western Reserve University,
10900 Euclid Ave, Cleveland, OH 44106}

\author{George H. Jacoby}
\email{jacoby@wiyn.org} \affil{WIYN Observatory\altaffilmark{2}, P.O. Box
26732, Tucson AZ 85726}

\and
\author{Robin Ciardullo}
\email{rbc@astro.psu.edu}
\affil{Department of Astronomy and Astrophysics,
Penn State University, 525 Davey Lab, University Park, PA 16802}

\medskip

\slugcomment{accepted for publication in the Astrophysical Journal}

\altaffiltext{1}{Research Corporation Cottrell Scholar and NSF CAREER Fellow}

\altaffiltext{2}{The WIYN Observatory is a joint facility of the University of
Wisconsin-Madison, Indiana University, Yale University, and the National
Optical Astronomy Observatory.}

\begin{abstract}
We report the results of a radial velocity survey of planetary nebulae (PNe)
located in the tidal features of the well-known interacting system NGC~5194/95
(M51).  We find clear kinematic evidence that M51's northwestern tidal debris
consists of two discrete structures which overlap in projection -- NGC~5195's
own tidal tail, and diffuse material stripped from NGC~5194.  We compare these
kinematic data to a new numerical simulation of the M51 system, and show that
the data are consistent with the classic ``single passage'' model for the
encounter, with a parabolic satellite trajectory and a 2:1 mass ratio.   We
also comment on the spectra of two unusual objects: a high-velocity PN which
may be associated with NGC~5194's halo, and a possible interloping
high-redshift galaxy.

\end{abstract}


\section{Introduction}

The ``Whirlpool Galaxy,'' M51 (NGC~5194, 5195)\footnote{Throughout this paper,
we use the designation M51 when considering both galaxies as a complete system;
when referencing the system's individual components, we use NGC~5194 and
NGC~5195.} is probably the most famous of all interacting galaxy systems.  As a
nearby system \citep[$d = 8.4 \pm 0.6$~Mpc;][hereafter FCJ]{fcj} with
grand-design spiral morphology (Rosse 1845; see Rosse 1880), distorted outer
isophotes \citep{zwicky59, burk78}, and an apparent bridge-like feature between
the primary, NGC~5194, and the secondary, NGC~5195, M51 has been extensively
studied as an example of tidally induced spiral structure
\citep[\eg][]{tully74, scoville83, rots90, zaritsky93}. In fact, the wealth and
detail of the observational data has made M51 a favorite target for dynamical
modeling, starting with the seminal work of \citet{tt72}. In the \citet{tt72}
study, many of the tidal features of the M51 system were explained by a
parabolic encounter of two galaxies with mass ratio of 3:1 viewed shortly after
the initial collision. Since then, in response to the ever-increasing amount of
observational data on the system, a number of alternate scenarios have been
proposed \citep{toomre78, howard1990, hern90, barnes98, salo00}.

Despite M51's long history of dynamical modeling, significant uncertainties in
the basic description of the system remain.  While the original \citet{tt72}
study proposed a very recent ($\sim 100$ Myr) collision, the discovery of M51's
long H~I tidal tail \citep{rots90} shifted the preferred solution to somewhat
later times (several hundred Myr past the initial collision) in order to give
the tail more time to develop \citep{hern90}.  More recently, \citet{salo00}
have suggested that a multiple passage model might be more appropriate for the
system.  Such a scenario appears to do a better job of explaining NGC 5194's
H~I velocity field, although the predicted structure for the H~I tidal tail is
more complex than is observed.

Furthermore, the simulations of \citet{salo00} used rigid halo models, which do
not self-consistently follow the orbital evolution of the system. Because the
multiple passage model relies on orbital decay to provide the proper second
passage, the lack of a self-consistent solution remains a concern for these
models.  As a result, these models have not followed the full dynamical
response of the system \citep{salo00}.  Consequently, no single scenario
satisfactorily explains all of the system's observational data \citep[see the
discussion in][]{barnes98}.

One reason for the continuing uncertainty about the M51 system is the lack of
kinematic information for the companion galaxy.  Unlike NGC~5194, NGC~5195
contains no neutral hydrogen, so the only kinematic data we have on NGC~5195
comes from measurements of the stellar kinematics of the system's inner disk
\citep{schweizer77}.  Since tidal kinematics provide strong constraints for
dynamical models of interacting galaxies \citep[\eg][]{hibbard95}, this dearth
of information at large radii is a significant stumbling block for unraveling
the evolutionary history of the system.

In principle, there is another dynamical tracer which can reveal the kinematic
structure of M51's tidal debris --- planetary nebulae (PNe).  Since PNe are a
normal and common phase of stellar evolution, their spatial distribution and
kinematics closely follows that of the stellar component as a whole.   As a
result, surveys for PNe can trace the distribution of stars to lower surface
densities than is possible with diffuse light.  More importantly, PNe are
extremely luminous emission-line sources.  At the distance of M51 ($d=8.4 \pm
0.6$~Mpc; FCJ), PNe surveys with 4-m class telescopes can reach $\sim 2$~mag
down the planetary nebula luminosity function, and the velocity of each PN can
be measured to $\sim 10$~\kms\ accuracy.  This makes PNe uniquely useful as a
kinematical probe of diffuse tidal structures, such as those found in the M51
system.

In 1997, FCJ surveyed M51 for planetary nebulae in order to obtain a distance
to the system via the planetary nebula luminosity function. This survey found a
substantial number of PNe directly west-southwest of NGC~5195 in a tidal
tail-like structure.  At first, this discovery was a bit of a surprise, since
the deep broadband images of \citet{burk78} place the western tail of NGC 5195
more to the northwest, and not at the location of these planetaries. However,
numerical models \citep[\eg Toomre 1994, as reported by][]{barnes98, salo00}
do predict the presence of tidal material from NGC~5194 in the region where
the planetaries are located.

In order to study this feature in more detail, and to provide kinematic data on
the diffuse tidal structures surrounding M51, we have conducted a radial
velocity survey of a significant fraction of the FCJ planetary nebula sample.
The ultimate goal of these observations is to provide a more complete
description of the kinematics of the M51 system, and test whether the
kinematics of NGC~5195's tidal features are consistent with the extant models.
Interestingly, our PN velocities reveal significant kinematic substructure in
the diffuse material to the west of NGC~5195; this fact, combined with
differences in the spatial distribution of the region's PNe and diffuse light,
implies that the observed tidal tail consists of two distinct but overlapping
features. We interpret these data in the light of published models and our own
new N-body model.

The outline of our paper is as follows: in \S2 we detail the original imaging
and follow-up spectroscopic observations.  In \S3, we describe our data
reduction and the determination of the planetary nebula velocities.  In \S 4
and 5, we describe the kinematic structure of M51's western tidal tail, and
compare the data to a new simulation which follows the response of both
NGC~5194 and its companion.   In \S 6, we describe two unusual objects whose
properties are significantly different from the bulk of the planetaries, and
discuss their implications. Finally, in \S7, we summarize our results.

\section{Observations}

The detection of M51's planetary nebula candidates was reported in
full in FCJ: here we briefly summarize their results.  The original
data were taken with the KPNO 4~m telescope and the T2KB detector. PN
candidates were defined as point sources detected manually that were
present on co-added images taken through a redshifted [O~III] $\lambda
5007$ filter (central wavelength 5017~\AA, full-width-half-maximum
31~\AA), but not present on similarly co-added frames taken through a
broader (275~\AA\ FWHM) off-band filter ($\lambda_c = 5300$~\AA).
Thus, at M51's systemic velocity \citep[463~\kms ;][]{schweizer77},
the survey was sensitive to emission-line objects with $-750 < v <
~1100$~\kms.  In addition, all PN candidates had to be completely
invisible on complementary $R$-band and H$\alpha$ images; this
requirement helped discriminate true PNe from compact H~II regions and
supernova remnants.  A total of 64 PN candidates were detected in the
survey; the location of these objects are displayed in the left panel
of Figure~1.  As the figure indicators, 45 of the candidate PNe
project onto a tail-like structure to the west of NGC~5195.  These PNe
do not simply follow the low-surface brightness features seen in the
deep image of \citet{burk78}; the PNe extend much further west of the
galaxy.  To illustrate this effect, the right panel of Figure~1 shows
a logarithmic stretching of the FCJ off-band image, binned $5 \times
5$ pixels to enhance the faint tidal features.

Also of note in Figure 1 is the paucity of PN candidates in the
visible disk of NGC 5194.  As noted by FCJ, this is due to the high
surface brightness of the spiral arms, and the confusion caused by the
system's many line-emitting H~II regions and supernova remnants.
However, outside the disk, the FCJ survey is uniform and complete.
FCJ confirmed this by adding artificial stars to the M51 onband frame
and determining the completeness limit, where the rate of recovery begins
to decline.  The lack of PN candidates southeast and due west of NGC~5194
is real:  there is an absence of stellar material in these regions.

\subsection{Improved Astrometry of M51's Planetaries}

In FCJ, astrometry of M51's PN candidates was performed using the
positions of stars in the {\sl Hubble Space Telescope\/} Guide Star
Catalog \citep[GSC;][]{lasker}.  Unfortunately, in most cases, the CCD
images of these stars were saturated; this forced FCJ to derive their
PN coordinates using a set of secondary standards measured on digitized
images of the Palomar Sky Survey.  Although the formal rms error of
this procedure was $0\farcs 49$, a two-step process such as this is
clearly susceptible to systematic error.

To reduce this error and maximize the signal-to-noise of our spectroscopy, we
re-derived the astrometry for M51's PN candidates using the more precise and
much higher density USNO-A 2.0 astrometric catalog \citep{monet1996,monet1998}
and the FINDER astrometric package from IRAF\footnote{IRAF is distributed by
the National Optical Astronomy Observatory, which is operated by the
Association of Universities for Research in Astronomy, Inc., under cooperative
agreement with the National Science Foundation.}.  The new coordinates, as well
as the previously measured $m_{5007}$ magnitudes, are given in Table~1; the rms
error of our astrometry is $0\farcs 4$.  Note that our PN positions are
significantly different from those given by FCJ:  not only is there a
systematic offset between the GSC and USNO catalogs ($-3\farcs 7$ in right
ascension and +$0\farcs 3$ in declination), but there are also noticeable
position-dependent terms in the GSC-based astrometry.  We attribute this
difference to the very small number (13) of GSC stars available for the
original astrometric measurements, and the complex spatial distortions of the
old KPNO 4~m prime focus corrector \citep{jacoby1998}.

\subsection{WIYN Spectroscopic Observations}

On UT 16 May 2001 we used the WIYN 3.5~m telescope to measure the radial
velocities of 43 of M51's planetary nebula candidates.  The specific instrument
configuration consisted of the Hydra fiber positioner with $2\arcsec$ diameter
red-sensitive fibers, the WIYN's telescope's bench spectrograph, and a
600~lines~mm$^{-1}$ grating blazed at $10\fdg 1$ in first order.  This setup
yielded spectra covering the wavelength range between 4400~\AA\ and 7200~\AA\
at 2.8~\AA\ (168~\kms) resolution with a 1.4~\AA~pixel$^{-1}$ dispersion.

Our data were collected in two fiber setups, the first targeted for
29~PNe (plus 3 sky fibers), the second for 27~PNe (plus 3 sky fibers).
The exposure times for these setups were 90~min ($3 \times 30$~min)
and 120~min ($4 \times 30$~min), respectively.  To provide a check on
the repeatability of our measurements, 14 PN candidates were observed
twice through different fibers.  Five candidate PNe were not detected
in our survey; we expect these were likely missed due to fiber
positioning errors, poor astrometry, or low fiber
throughput\footnote{While it is strictly possible that some of these
candidates were spurious detections (eg. cosmic rays), all candidates
from FCJ were visually inspected and confirmed to be stellar, making
this hypothesis unlikely.}.  This left us with a total of 37 objects
for further analysis.

The wavelength calibration for each spectrum was provided by a CuAr comparison
arc taken immediately before each observation.  In addition, to test our
ability to centroid the weak lines of M51 PNe, we also obtained a set of CuAr
arcs at two different exposures, 10~s and 60~s.  By comparing measurements of
weak arc lines (\ie\ lines comparable in strength to the [O~III] $\lambda 5007$
emission feature of M51 PNe) to measurements of the same lines taken at six
times the exposure level, we placed a limit on the internal errors associated
with our velocity measurements.  Our centroiding uncertainty was typically 0.1
to 0.15 pixels, or less than 0.1 of a resolution element.  This implies an
expected velocity error for our measurements of $< 20$~\kms.

\section{Data Reduction and Analysis}

The individual Hydra images were first pre-processed (bias subtracted and
flat-fielded) with the DOHYDRA routine within IRAF, and then averaged together
with SCOMBINE{}.  A dispersion relation for each spectrum was then obtained by
fitting 14 bright spectral lines on the CuAr arc spectrum with a 4-th order
spline; the resulting solution had a 0.02~\AA\ rms error. Once this was done,
the spectra of our three sky fibers were averaged and then subtracted from
those of our program objects to create a sky-subtracted spectrum for each
planetary.  Finally, the spectra of the 14 objects observed with both setups
were combined to increase their signal-to-noise.

After completing these reduction procedures, we examined each spectrum for the
presence of an unresolved emission line at the approximate location of [O~III]
$\lambda 5007$ at the redshift of the galaxies.  Of the 42 PN candidates
observed, 36 had this feature, and $\sim 23$ also showed evidence for the
weaker [O~III] line at $\lambda 4959$.  In addition, one PN candidate (observed
with both setups) exhibited a single, broad ($\sim 350$~\kms\ FWHM) emission
feature; this object is probably a background object and is discussed in \S
6.1. Figure~2 displays the PN spectra in the wavelength region about [O~III]
$\lambda 5007$.

Velocities were determined for the PN candidates by assuming the
unresolved emission line near 5015~\AA\ is, indeed, redshifted [O~III]
$\lambda 5007$, and centroiding the line with the SPLOT routine in
IRAF{}.  To avoid degrading the signal-to-noise, these measurements were
carried out on the spectra prior to sky subtraction; this makes
absolutely no difference to the analysis, as the sky around 5000~\AA\
in our spectra is completely negligible, and velocities derived before
and after sky subtraction showed no measurable difference.  The PN
velocities, corrected for the Earth's motion about the Sun, are listed
in Table~1.  Table~1 also notes those PN candidates that were not
detected in our survey.  Note that the tabulated velocities are based
entirely on the emission line at [O~III] $\lambda 5007$.  Although the
weaker lines of [O~III] $\lambda 4959$ and H$\alpha$ were detected in
a number of spectra, the signal-to-noise of these lines were too low
for an accurate velocity measurement.

To confirm our identification of the $\lambda 5015$ emission line as
redshifted [O~III] $\lambda 5007$, we used the velocities of Table~1
to shift each spectrum back into its rest frame.  We then scaled the
spectra so that all the putative $\lambda 5007$ lines had the same
weight, and summed the data to create one single composite spectrum.
This spectrum is displayed in Figure~3.  The left-hand panel of the
figure clearly shows both [O~III] $\lambda 5007$ and [O~III] $\lambda
4959$.  Despite the fact that the system throughput is decreasing
rapidly shortward of 5000~\AA, the ratio of the two oxygen lines
(3.2:1) is nearly identical to that expected from atomic physics
\citep[2.98;][]{sz00}. This ratio confirms that virtually all the observed
objects are emission line objects associated with M51 (see Freeman
\etal\ 2000 for a similar analysis on Virgo cluster PN spectra).  The
presence of H$\alpha$ and [N~II] $\lambda\lambda 6548,6584$ in the
right hand panel of Figure~3 provides additional evidence for this
interpretation.  The relatively strong [N~II] to H$\alpha$ ratio
($\sim$ 0.7) may suggest a fairly high abundance in the stars
associated with the PNe in M51, as was seen in M32 \citep{srm98}.

To estimate the random uncertainty associated with our velocity
measurements, we used the sub-sample of PNe observed with both Hydra
setups.  Fourteen PN were observed twice.  Of these, 12 had their
[O~III] $\lambda 5007$ emission detected in both setups, one was
detected only in Setup~2, and one had a broad, asymmetrical emission
line that was unsuitable for the experiment.  A comparison of the
velocities derived from the individual spectra of the 12 objects
appears in Table~2.  In general, the agreement between the two setups
is excellent: if we ignore the highly discrepant velocity difference
for PN 11, the mean difference between Setup~1 and Setup~2 is $0 \pm
4$~\kms, and the dispersion is 12~\kms.  Thus the velocity error in a
single spectrum is 8 \kms\ if the measurements are completely
uncorrelated; however, we expect this is not quite the case, and the
true error will lie between 8 \kms\ and 12 \kms.  We have adopted the
upper limit of 12 \kms\ as the measurement error of our derived
velocities.

Figure~4 shows the locations of the spectroscopically-measured PNe
overlaid on the [O~III] image of FCJ{}.  The PN velocities are denoted
by color.  The most striking feature of the figure is the velocity
structure of the low surface-brightness tail just west of NGC~5195.
As the histogram of Figure~5 indicates, the distribution is distinctly
bimodal.  Two thirds of the PNe in this region have radial velocities
greater than the systemic velocity of either galaxy in the interaction
\citep[463~\kms\ for NGC~5194; 600~\kms\ for
NGC~5195;][]{schweizer77}.  Conversely, the velocities of the
remaining PNe are at or below systemic velocity of NGC 5194.  This
bimodal behavior contrasts with the broad but generally unimodal
velocity distribution exhibited by the PNe east of NGC~5195 and in the
outer disk of NGC~5194; these objects generally have velocities
somewhere between the two systemic velocities.  Since no H~I is
present in the western tidal tail of M51, the bimodal distribution of
Figure~5 is the first measurement of velocity substructure in the
region.

\section{The Kinematic Structure of the Western Tail}

The velocity field of the PNe of M51 is complex.  This is particularly
apparent in the western tail of NGC~5195, where one sees a multi-modal
distribution of velocities.  Figure~6 illustrates the velocity
structure by displaying the distribution of PN velocities north of
NGC~5194, along an east-west swath across the face of NGC~5195. Two
things stand out in this figure. First, the velocity spread at any
position is highly clumped; this implies that we are not simply
viewing a single, kinematically hot component (\ie\ a galaxy halo).
Instead the data suggest the existence of multiple kinematically cold
components which likely arise from different parent populations.  The
second feature of Figure~6 is the presence of a significant gradient
in the PN velocities. Since this gradient is in the same sense as that
observed for the stars in the inner disk of NGC~5195
\citep{schweizer77}, the data suggest that the {\it bulk\/} of the
observable PNe in this region come from NGC~5195 itself, rather than
being captured or tidal debris from NGC~5194.

Comparing the spatial distribution and kinematics of M51's PNe to
published dynamical models is somewhat difficult, because these
studies either did not model the detailed response of the companion
\citep[\eg][hereafter H90]{hern90} or employed simple N-body models
which did not capture the full dynamical evolution of the system
\citep{toomre78, salo00}.  However, we can make some qualitative
inferences based on the published models. Two broad classes of models
exist: single passage models, like the original \citet{tt72} model
(later refined by H90), and multiple encounter models, like the one
recently proposed by \citet[][hereafter SL00]{salo00}.  Both models
can reproduce the overall morphology of the M51 system fairly well;
the multiple passage model better describes some of the detailed HI
kinematics of NGC 5194, but predicts a more complicated structure for
the long HI tidal tail than is actually observed.

In terms of the PN population west of NGC~5195, both classes of models
predict that material from NGC~5194's disk will be dispersed in this
region.\footnote{While this material was not evident in the H90
simulations, that is likely due to the relatively small particle
numbers used in the study.}  In the multiple passage model, this
material has been perturbed in the very recent ($\sim 10^8$~yr) past
by the second passage of NGC~5195.  As a result, the model predicts a
higher radial velocity for the material than is produced by
corresponding single passage models (for example, see Figure~4 in
SL00).  In second passage models, the velocities of the region range
from 400 to 525~\kms, while in single passage models, the velocities
are between 385 and 425~\kms.

Since the PN velocities west of NGC~5195 (315-480~\kms) lie between
these two ranges, we cannot reliably discriminate between the two
models.  There is some hint of a blueward gradient towards the west in
our PN data, and this is more consistent with the SL00 multiple
passage model than a single passage model, but the small number of
velocities makes this result tentative at best.  Furthermore, it
should be re-emphasized that the PN kinematics in the western tail of
the system are truly bimodal: there is a complete 140~\kms\ gap
between two kinematic components.  While SL00 did not show the
kinematic structure of the companion's tidal debris, it is hard to
understand how the galaxy could undergo several close passages with
NGC~5194 and still have its material exhibit such discrete segregation
in velocity space.

\section{Modeling the Kinematics of the Planetary Nebulae}

Our PN observations do not explicitly rule out either class of model, but, as
SL00 point out, there are a number of features which argue for a multiple
passage model for M51.  First, the kinematics of NGC~5194's H~I tail and the
apparent tidal bridge between NGC~5194 and 5195 are better fit by a multiple
passage model than a single passage model.  Second, single passage models imply
that a long time has elapsed since perigalacticon; it is not clear whether the
strength of NGC~5195's tidal features is consistent with this prediction.
Finally, many single passage models have trouble reproducing the morphology of
NGC~5194's northwestern tidal plume (which gives rise to the blueshifted PNe
component in the system's western tail). For example, in the SL00 models, this
tail extends to the northwest of NGC~5195 (see their Figure~1), while in the
models of Toomre (1994, as reported by Barnes 1998), the tail has the correct
orientation, but the mass
ratio of the galaxy pair is a hefty 1:1.  These various inconsistencies make it
interesting to re-investigate the single passage scenario to see if we can
construct a model that better matches the spatial and kinematical distribution
of the tidal debris. We can do this by revisiting the H90 single passage
simulation and explicitly including the response of the companion.

We began the simulation process by first defining the structure of each
galaxy.  The galaxies were constructed in the manner described by
\citet{hern93}, and consisted of an exponential stellar disk, a central
bulge, and a truncated isothermal halo. The mass ratio of the components
was $M_d:M_b:M_h = 1:0.33:5.8$.  The companion was identical to the
primary, except scaled down in mass by a factor of $f= 1/2$.  The size and
circular velocity of the model was scaled by $f^{1\over 2}$ and $f^{1\over
4}$ respectively, which preserved disk surface density between the two
models and provided a Tully-Fisher- like $M \sim V_c^4$ scaling of the
model. The simulations were then evolved in a self-consistent manner using
TREECODE \citep{hern87}, using a total of $N=55,296$ particles for the
models which survey parameter space, and with a total of $N=425,984$
particles ($N_{disk}$=131,072; $N_{bulge}$=16,384; $N_{halo}$=65,536 in
each galaxy) for the final model.  The simulations surveying parameter
space were run using a parallel processing Beowulf cluster.

To begin the interaction modeling, we used the parameters of the self-
consistent simulation of H90 as our starting point. These parameters include an
NGC~5194/NGC~5195 mass ratio of 2:1, an initially parabolic orbit with a
perigalacticon distance of 6 disk scale lengths, and a disk geometry for
NGC~5194 of ($i_1$,$\omega_1$)= ($-$70,$-$45), where $i$ and $\omega$ are the
orientation angles as defined by \citet{tt72}.  After surveying simulations
with varying $\omega_1$ we settled on a value of $\omega_1=0$, which resulted
in a longer and thinner tidal tail for NGC~5194 (as perigalacticon occurs in
NGC~5194's disk plane) and kept the companion closer to NGC~5194 in projection
while still preserving a fairly high projected velocity difference between the
two galaxies.

With the parameters of NGC~5194's disk and relative orbit set in this
manner, we next ran a grid of simulations covering the range of possible
orientations $(i_2,\omega_2)$ for NGC~5195.  To quickly narrow down the
list of possible geometries, we examined the simulations to select those
which gave the closest morphological match to the M51 system.  In doing
this, we paid particular attention to the tidal plumes northeast and
southwest of NGC~5195.  We were constrained in viewing geometry by the
need to match the projected position and relative velocity  of the galaxy
pair, and by the observed $20^{\circ}$~inclination of NGC~5194's disk
\citep{tully74}.  NGC~5195's observed inclination also provided a
constraint to the simulations.  Based on the ellipticity of NGC~5195's
isophotes, \citet{schweizer77} inferred an inclination angle of $48^\circ$
for the system.  Although this estimate may be affected by tidal deformation, the
galaxy's appreciable velocity gradient \citep[$\sim 120$~\kms;][]{schweizer77}
requires that it not be too face-on.  The final constraint on the models
came from NGC~5195's rotation, which is east to west in the inner portions
of the galaxy \citep{schweizer77}.  The models which best matched these
morphological and kinematic requirements had geometries for the companion
in the range of $(i_2,\omega_2)= (105 \pm 15, -30 \pm 15)$, with our
favored model having $(i_2,\omega_2)= (110, -30)$.

An evolutionary sequence of our model is shown in Figure~7. The
initial passage occurred 280 Myr ago, similar to H90's best match time
of $\sim$ 300 Myrs past perigalacticon. At the present time (T=0), the
galaxies are widely separated ($\Delta R \sim$ 50 kpc) and moving
apart from one another quite rapidly ($\Delta V \sim$ 150 km/s; in
other words, since NGC 5195 is behind NGC 5194 in our model, the
observed radial velocity difference closely reflects the separation
velocity). This is quite unlike the simulation of SL00, which favors a
much closer true separation and an orbit which has carried the
companion just past a second disk passage.

The model above plausibly reproduces the main morphological and kinematic
properties of the M51 system, including the sharp northwestern tidal tail of
NGC~5195. We now compare the tidal kinematics in the model against the
kinematics of the planetary nebulae. To do this, we examined the projected
velocity distribution of the model in the four regions of the system which
contain the majority of our planetary nebulae. These regions are illustrated in
Figure~8. Note that a strict comparison between the PNe and the model is not
possible, since the selection criteria for the two datasets are very different:
while the simulation samples velocities weighted by mass, the observational
data suffer from constraints imposed by our non-random target selection and
fiber crowding. Therefore, in order to compare the model to the PN velocities,
our approach was to look more for similar patterns of kinematic substructure,
rather than to perform a detailed one-to-one matching.

Field~1, which is immediately west of NGC~5195, has the most interesting
velocity field.  The stellar kinematics in this region are distinctly bimodal:
there is a velocity peak at $v_r \sim 250$~\kms\ which corresponds to the tidal
plume from NGC~5195, and a second peak around $v_r \sim -50$~\kms, which comes
from material pulled out of NGC~5194.  This echoes the distribution of PN
velocities, and explains the existence of three blue-shifted PNe (relative to
NGC~5195 systemic) just west of the body of NGC~5195: it is here that
NGC~5194's tidal plume projects across the face of NGC 5195. The three other
blue-shifted PNe further west and south of NGC~5195 are either with NGC~5194's
plume or with the diffuse extension of the NGC~5194's disk. In either case,
their velocities more closely reflect NGC~5194's systemic velocity rather than
that of NGC 5195's tail.  The model also has a clear gap in velocity between
the two components; this results from the fact that these two tidal features
are spatially distinct and kinematically cold.

Field~2 is east of the apparent tidal bridge which links NGC~5194 and NGC~5195.
Like Field~1, this region, too, shows a multimodal velocity pattern, which
arises from the discrete populations of NGC~5194's bridge and NGC~5195's
eastern tidal tail.  Unfortunately, our model for Field~2 does not match the PN
velocity distribution very well:  although there is a cluster of PNe near the
predicted velocity of NGC~5195's tail, there are no blue-shifted PNe in our
sample. Instead, the region contains two unexplained red-shifted PNe.  This
discrepancy is somewhat disturbing, but it hard to make a strong statement
about the quality of the agreement based on only 5 objects.  This is
particularly true since our models do not account for the presence of
kinematically hot populations (see \S 6.2 for further discussion on this
point).

Fields~3 and 4 are much more straightforward to interpret. Field~3 covers the
southwestern portion of NGC~5194's disk, and Field~4 includes the northeastern
portion of NGC~5195. In both fields, the model velocity distribution is
single-peaked, and representative of the large scale velocity field of the
superposed galaxy. The match with the PN velocities in Field 3 is excellent:
both the velocity mean and the dispersion agree with the predictions of the
model.  The agreement of Field~4 is slightly worse, since the PNe there have
slightly larger velocities than predicted.   Of course, the prior constraints
on the model -- that the southern portion of M51 and the eastern portion of
NGC~5195 both be blueshifted relative to their systemic velocities --
practically ensures that the match between the model and the mean of the PN
velocity distribution is close.  However, the velocity dispersion of each field
is also reproduced well by the model, which is not guaranteed by these velocity
constraints.

While our model reproduces many of the morphological and kinematic features of
the M51 system, there are still some discrepancies.  Morphologically,
NGC~5195's eastern tail seems to tilt slightly more south than it does in our
model, and the apparent bridge between the modeled galaxies shows more
curvature than is actually observed.   Also, as the data of Field~1 and 4
indicate, our model predicts too large of a velocity gradient across NGC~5195.
There are a number of ways to fix this problem, such as changing the
orientation of NGC~5195's disk ($i_2, \omega_2$) so that the system is observed
more face-on, or reducing NGC~5195's mass (\ie\ by using a 3:1 mass ratio for
the system) and thus decreasing the galaxy's rotation velocity.  However, our
main goal here is to see whether a classic single passage model can reproduce
the global morphology and kinematics of the tidal features of M51; clearly,
this model presents such a solution.

Nevertheless, we leave with a cautionary note. Parameter space is wide, and the
parameters describing the interaction are all tightly coupled. The morphology
of tidal debris depends on the encounter geometry, the orbital parameters, the
viewing angle and time, and the structural properties of the galaxies. For
example, the dark matter distributions of NGC~5194 and NGC~5195 at large radius
have a profound impact on the orbital evolution of the system.  Since the
viewing angle is constrained by the projected separation and velocity of the
galaxy pair, the best-fitting value for this parameter will change if the dark
matter distributions are modified.   A different viewing angle will then alter
the observed morphology of the system and the projected distribution of the
tidal debris. Because of this degenerate coupling of parameters, iterating one
parameter while holding the others fixed can be a misleading exercise and lead
one to a ``local minimum'' in parameter space that ignores the much wider
parameter space available.  Simulations such as the one presented here should
be viewed as caricatures of the real system, with broad uncertainties and
lingering concerns about uniqueness.

With this in mind, we have no illusions that our model is the last word on the
M51 solution.  We have focused mainly on a comparison of the morphology, global
kinematics, and PNe velocity field.   Further revisions -- and perhaps very
different solutions -- may be possible based on more detailed matching of the
velocity field.  Indeed, the multiple passage models of SL00 argue for a more
tightly bound ``second passage'' encounter based on the detailed HI kinematics
of the system, and may do an equally good job of modeling the PNe kinematics.
However, it is unclear whether these second passage models can reproduce the
bimodality {\it and} kinematic coldness of the distinct tidal features which
comprise the western tail.  Self- consistent simulations and detailed
kinematics for these second passage models will be useful in testing these
models; unfortunately, the uniqueness of these solutions will be even more
questionable since they rely on the interplay between the (largely
unconstrained) dark matter halos, dynamical friction, and orbital decay to
obtain the best match for the M51 system.

\section{Unusual Objects}

There are two PNe candidates, PN~52 and PN~27 that are in clear disagreement
with the dynamical model we have adopted.  We now discuss these objects and
attempt to determine their true nature.

\subsection{PN 52 - A Ly$\alpha$ galaxy?}

Figure~9 compares the optical spectrum of PN~52 to that of another PN
in the survey.  Photometrically, both objects are similar, with
$\lambda 5007$ fluxes of $\sim 1.1 \times
10^{-16}$~ergs~cm$^{-2}$~s$^{-1}$.  However, as the figure
illustrates, the two sources are different spectroscopically.  First,
PN~52 has no detectable [O~III] $\lambda 4959$ line -- true planetary
nebulae must have this line, and, as the comparison spectrum
demonstrates, this line should have been seen.  It was
not\footnote{Although there is a hint of emission near $\lambda 4959$,
its centroid is 4~\AA\ too blue for it to be the [O~III] line.}, and,
in fact, no other lines have been reliably detected in this spectrum.
Second, the line profile of PN~52 is anomalously broad and asymmetric.
The width of the object's lone emission line is approximately 6~\AA\
(FWHM); for comparison, the spectrograph's instrumental profile is
3.0~\AA (FWHM).  No other PN in the sample has a resolved
line. Finally, if we identify the emission line as [O~III] $\lambda
5007$, then the heliocentric velocity of the object is $-176$~\kms, or
$-640$~\kms\ systemic. Such an object cannot be gravitationally bound
to M51.

The most likely explanation for PN~52 is that it is not a planetary
nebula at all, but a higher redshift source whose emission line has
been red-shifted into FCJ's [O~III] $\lambda 5007$ filter.  This is
quite plausible: in a deep blank-field survey far away from any galaxy
or cluster, \citet{blank} found a population of extragalactic sources
that mimic the properties of planetary nebulae.  If we scale the
results of this survey to the area and depth of the M51 survey
(256~arcmin$^2$, $m_{lim} = 26.3$) and the width of the narrow band
filter, we should expect $0.8 \pm 0.5$ contaminating sources to be
present in the FCJ sample.

PN~52 is not the first contaminating source to be found in a deep
[O~III] $\lambda 5007$ planetary nebula survey.  Some intracluster PN
candidates in Virgo have turned out to be high-redshift objects
\citep{kud00,freeman00}.  Also, in a recent PN survey of NGC~4697,
\citet{mendez01} reported that three of his 535 PN candidates were
interlopers.  Based on these data, it is clear that some amount
of contamination will be present in all [O~III] $\lambda 5007$
observations fainter than $m_{5007} = 26.0$.  Future radial velocity
surveys of PNe in external galaxies need to take this effect into
account.

What is the true identity of PN~52?  In order to be listed as a PN
candidate by FCJ, the emission equivalent width
must be extremely large, $\gtrsim 100$~\AA.  This, along with the
object's asymmetric line profile, suggests that the source is a
Ly$\alpha$ galaxy at redshift $z \sim 3.12$.  We caution that this
interpretation is not conclusive: \citet{stern} have shown that lower
redshift objects can masquerade as high-$z$ Ly$\alpha$ emitters.  To
conclusively confirm the identity of this object, deeper spectroscopy
is needed, to either detect the Lyman decrement or find additional
emission lines in the spectrum.

\subsection{PN 27 - A highly blueshifted PN}

Spectroscopically, PN~27 is very similar to the other PNe surveyed in this
program: its [O~III] $\lambda 5007$ emission line is strong and unresolved, and
the weaker [O~III] $\lambda 4959$ line is present at roughly $1/3$ the strength
of $\lambda 5007$.  However, in terms of location and motion, the object is
unique.  PN~27 sits SSE of M51 well away from any extended tidal plume (see
Figure~1, left), and it has a radial velocity of 194~\kms.  In other words,
PN~27 is blue-shifted by 269~\kms\ relative to NGC~5194, and 406~\kms\ relative
to NGC~5195.  This remarkable velocity is not due to measurement error.  PN~27
was observed twice: in both cases, both lines of [O~III] were detected.   The
velocity difference between the two measurements is only 3~\kms.

If PN~27 belongs to NGC~5194, then it is moving at $-269$~\kms\ with respect to
NGC~5194.  What can cause this motion?  One possibility is that PN~27 belongs
to the stellar halo of the galaxy.  In the Milky Way, such high-velocity
objects do exist, as the vertical velocity dispersion in the spheroidal
component is $\gtrsim 100$~\kms\ \citep{mb}.  If the stellar halo of NGC~5194
is similar to that of the Milky Way, then it is at least conceivable that a
high velocity PNe could have been observed.

We can estimate how likely the halo PN explanation is by using the bulge/disk
parameters for NGC~5194 given by \citet{baggett}.  As column (4) of Table~1
demonstrates, the FCJ survey of NGC~5194 was restricted to galactocentric radii
$2\farcm 8 \lesssim R_{\rm iso} \lesssim 8\farcm 0$. If we extrapolate the
Baggett \etal\ $r^{1/4}$-law to these radii, adopt the FCJ distance of $\sim
8.4$~Mpc, and apply the bolometric correction of an old population
\citep[$-0.8$;][]{p5}, then we obtain a value of $L_{\rm bol} \sim 2 \times
10^9 L_\odot$ for the amount of spheroid (bulge$+$halo) luminosity contained in
the FCJ survey field.  To translate this into an expected number of PNe, we
assume old stellar populations produce $\sim 20 \times 10^{-9}$~PNe (within
2.5~mag of the luminosity function cutoff, $M^*$) per unit bolometric
luminosity \citep{c95}.  When we scale this to the FCJ survey depth (1.2~mag
down the PN luminosity function), we get a value of $\sim 14$ for the number of
spheroid PNe that could have been detected.  Since we obtained spectroscopy for
roughly half of the FCJ planetary nebula candidates, the result is that we
might expect $\sim 7$~spheroid PNe to be present in our sample.

Of course, not all halo objects have large vertical velocities.  For example,
of the 95 probable Milky Way halo RR Lyrae stars studied by \citet{layden},
only two have a vertical velocity as great as 269~\kms.  If the spheroidal
component of M51 is similar to that of the Milky Way, then the probability of
finding a high-velocity PN in our sample is small, $\sim 0.1$.  This number is
highly uncertain, as it is derived from a rather large photometric
extrapolation and an assumed `maximal' $r^{1/4}$ spheroid.  Nevertheless, it
does demonstrate that the presence of halo PNe in our sample is
possible.\footnote{As suggested in \S 5, a small number of our PN have
velocities that do not agree with the model predictions.  It is possible that
these objects are also halo PNe.}  Such a `hot' population was not explicitly
included in our models.

An alternative explanation for the high velocity of PN~27 is that it is a
`runaway star', where the ejected planetary obtained a high velocity from
either the tidal interaction \citep[\eg][]{byrd1979}, or from some other
dynamical process.  However, we believe this hypothesis to be less likely --
our dynamical models show no evidence for an extremely blueshifted population,
and none of the hypothesized models for high-velocity neutron stars
\citep[\eg][]{cc98}, or O \& B stars seem relevant in this case.

The best way to distinguish between these two scenarios is through deeper PN
spectroscopy.  If the halo hypothesis is correct, we might expect PN~27 (and a
few other objects) to have a sub-solar oxygen abundance and perhaps enhanced
ratios of C/O and N/O; such properties have been observed for halo PNe in the
Milky Way \citep{how97}.  Deeper spectroscopy is also desirable for an improved
kinematic model of the system.  If the metallicities of the PNe could be
derived, this information could be used to constrain the place of origin of
individual objects.

\section{Summary}

We have presented the velocities of 36 PNe in the interacting galaxy system
M51.  The planetaries have a velocity structure that is clearly multi-modal and
complex, especially in the tidal debris west of the companion galaxy NGC~5195.
In this region, we find two distinct kinematic components, consistent with the
idea that we are viewing two tidal features -- one from each galaxy --
cospatially in projection.  Such features have been predicted by numerical
simulations \citep[\eg][]{salo00}, but the component from NGC~5194
has remained undetected until now.  The PN kinematics do not yet discriminate
between the various dynamical scenarios for the M51 system: both single- and
multiple-passage models predict multiple kinematic features in this region. We
do, however, note that the kinematic {\it coldness\/} of these features may be
difficult to reproduce in a multiple passage scenario, where the galaxies have
experienced several perturbative events. Self-consistent dynamical models of
the multiple passage scenario are needed to test this effect.  We expand on the
previous simulation of \citet{hern90} and present a new single-passage model
that reasonably reproduces the morphology and kinematics of the system's tidal
debris.

How can we better constrain the dynamical evolution of the M51 system?
Observationally, the FCJ survey of M51 probed only the first 1.2~mag of the
planetary nebulae luminosity function.  The sample of 64~PNe found in this
survey could easily be increased by a factor of $\sim 3$ by going deeper with
current 4~m and 8~m class telescopes.   There are roughly 10 PNe in an 18
arcmin$^{-2}$ area in the FCJ sample of the northwestern tidal tail.   Assuming
roughly $\sim 20 \times 10^{-9}$~PNe per bolometric solar luminosity (2.5 mag
down the PNLF) for a stellar population \citep[\eg][]{c95}, and a distance of
8.4 Mpc, this PNe density translates to a stellar surface brightness of $\mu_V
\sim 24.5$ mag arcsec$^{-2}$.   Allowing for a 3-fold increase in the number of
PNe detected in deeper surveys, and adopting a (conservative) minimum of 10 PNe
in a given region to derive useful kinematic information, PNe could be used to
probe surface brightnesses as faint as $\mu_V\sim 26$ mag arcsec$^{-2}$.  The
kinematics of such low-surface brightness features cannot be measured in any
other way.

In addition to a deeper M51 PNe survey, more information on the secondary,
NGC~5195, would be extremely helpful.  In particular, integral-field
spectroscopy \citep[\eg][]{and1999} of the galaxy's stellar component could
dramatically improve our constraints on galactic rotation and inclination
angle.   Finally, better theoretical models are needed to predict the detailed
tidal kinematics probed by the PNe. This is particularly true for the
multiple-passage models, for which no fully self-consistent solution has yet
been developed.

\acknowledgments We thank the observing staff at the WIYN telescope,
especially Di Harmer, for assisting us in obtaining excellent data.  We
also thank Paul Harding for useful discussions.
This work was supported by the NSF through grants AST-9876143 (JCM) and AST
0071238 (RC) and through a Research Corporation Cottrell Scholarship (JCM).

\begin{deluxetable}{ccccccc}
\tablewidth{0pt} \tablenum{1} \tabletypesize{\footnotesize} \tablecaption{M51
Planetary Nebula Candidates} \tablehead{
&&&\colhead{$R_{\rm iso}$} &&\colhead{Velocity} \\
\colhead{ID} &\colhead{$\alpha$(2000)} & \colhead{$\delta$(2000)}
&\colhead{(arcmin)} &\colhead{m$_{5007}$} &\colhead{(\kms)}
&\colhead{Notes\tablenotemark{a}} } \startdata
 1 & 13 29 50.61 & +47 06 27.5 & 5.3  &25.15 &535 & S       \\
 2 & 13 30 03.73 & +47 17 57.3 & 6.5  &25.34 &657 & S       \\
 3 & 13 30 16.82 & +47 14 01.8 & 4.7  &25.35 &657 & S       \\
 4 & 13 29 38.52 & +47 17 38.2 & 6.4  &25.36 &724 & S,T     \\
 5 & 13 29 23.33 & +47 17 25.3 & 7.6  &25.38 &664 & S,T     \\
 6 & 13 29 44.86 & +47 09 15.0 & 2.8  &25.38 &522 &         \\
 7 & 13 29 29.37 & +47 15 11.1 & 5.3  &25.39 &406 & S,T     \\
 8 & 13 29 48.90 & +47 16 37.6 & 5.0  &25.42 &476 & S,T     \\
 9 & 13 30 14.89 & +47 14 43.4 & 4.8  &25.44 &500 & S       \\
10 & 13 29 48.95 & +47 17 21.6 & 5.7  &25.48 &432 & S,T     \\
11 & 13 29 57.33 & +47 17 29.0 & 5.9  &25.51 &704 & T       \\
12 & 13 29 38.99 & +47 08 05.0 & 4.3  &25.52 &488 &         \\
13 & 13 29 37.28 & +47 17 25.9 & 6.3  &25.54 &713 & S,T     \\
14 & 13 29 29.07 & +47 17 19.7 & 6.9  &25.55 &736 & S,T     \\
15 & 13 29 35.89 & +47 16 42.0 & 5.8  &25.55 &655 & S,T     \\
16 & 13 29 36.65 & +47 07 07.0 & 5.3  &25.62 &448 & S       \\
17 & 13 29 48.04 & +47 15 57.8 & 4.3  &25.62 &687 & S,T     \\
18 & 13 29 49.23 & +47 17 23.5 & 5.7  &25.62 &    & S,T     \\
19 & 13 30 00.56 & +47 17 08.3 & 5.6  &25.63 &\dots    & N,T     \\
20 & 13 29 50.36 & +47 17 11.9 & 5.5  &25.65 &    & S,T     \\
21 & 13 29 36.93 & +47 08 35.0 & 4.1  &25.66 &511 &         \\
22 & 13 29 34.49 & +47 17 38.9 & 6.7  &25.66 &    & S,T     \\
23 & 13 29 46.61 & +47 17 10.3 & 5.6  &25.68 &413 & T       \\
24 & 13 29 46.03 & +47 15 31.2 & 4.0  &25.68 &    & S,T     \\
25 & 13 29 31.22 & +47 15 47.2 & 5.5  &25.68 &430 & S,T     \\
26 & 13 29 30.08 & +47 16 31.3 & 6.2  &25.68 &629 & S,T     \\
27 & 13 30 05.70 & +47 07 19.4 & 4.9  &25.69 &194 & S       \\
28 & 13 29 42.63 & +47 18 18.5 & 6.8  &25.70 &745 & S,T     \\
29 & 13 29 40.21 & +47 15 12.8 & 4.1  &25.70 &692 & S,T     \\
30 & 13 29 40.95 & +47 09 05.8 & 3.3  &25.74 &576 &         \\
31 & 13 30 00.01 & +47 16 57.5 & 5.4  &25.77 &    & T       \\
32 & 13 29 42.68 & +47 18 52.2 & 7.4  &25.81 &\dots    & N,S,T   \\
33 & 13 30 01.11 & +47 04 48.9 & 7.1  &25.83 &478 & S       \\
34 & 13 29 55.15 & +47 18 29.1 & 6.8  &25.85 &    & S,T     \\
35 & 13 30 07.93 & +47 16 22.0 & 5.3  &25.86 &\dots    & N,T     \\
36 & 13 29 44.02 & +47 16 13.7 & 4.8  &25.88 &\dots    & N,S,T   \\
37 & 13 29 30.87 & +47 14 36.1 & 4.7  &25.90 &315 & S,T     \\
38 & 13 29 51.23 & +47 16 05.9 & 4.4  &25.93 &    & T       \\
39 & 13 29 45.83 & +47 17 29.0 & 5.9  &25.97 &    & T       \\
40 & 13 29 17.44 & +47 15 21.5 & 7.0  &25.98 &    & S,T     \\
41 & 13 30 05.40 & +47 17 36.1 & 6.3  &26.00 &    & S,T     \\
42 & 13 29 48.52 & +47 17 18.1 & 5.7  &26.00 &    & T       \\
43 & 13 30 12.65 & +47 14 08.0 & 4.2  &26.00 &508 & S       \\
44 & 13 29 46.53 & +47 06 54.4 & 4.9  &26.02 &534 & S       \\
45 & 13 29 43.93 & +47 15 48.6 & 4.4  &26.03 &    & T       \\
46 & 13 30 08.25 & +47 16 18.8 & 5.3  &26.03 &515 &         \\
47 & 13 29 50.41 & +47 05 38.7 & 6.1  &26.08 &    & S       \\
48 & 13 29 24.23 & +47 15 02.7 & 5.9  &26.08 &    & S,T     \\
49 & 13 29 53.17 & +47 05 44.9 & 6.0  &26.09 &    & S       \\
50 & 13 29 44.80 & +47 16 04.5 & 4.6  &26.15 &    & S,T     \\
51 & 13 29 51.33 & +47 18 33.7 & 6.9  &26.16 &692 & S,T     \\
52 & 13 29 16.27 & +47 13 00.7 & 6.3  &26.17 &--- & G,S,T   \\
53 & 13 29 51.12 & +47 17 19.7 & 5.7  &26.17 &    &         \\
54 & 13 29 55.45 & +47 18 57.7 & 7.3  &26.21 &    & T       \\
55 & 13 29 54.70 & +47 16 20.0 & 4.7  &26.24 &    & T       \\
56 & 13 29 23.58 & +47 17 13.6 & 7.4  &26.26 &    & S,T     \\
57 & 13 30 08.28 & +47 17 52.7 & 6.8  &26.28 &593 & S       \\
58 & 13 29 44.01 & +47 05 51.9 & 6.0  &26.29 &\dots    & N,S     \\
59 & 13 29 46.96 & +47 05 36.3 & 6.2  &26.35 &    &         \\
60 & 13 29 45.44 & +47 06 34.5 & 5.3  &26.37 &532 &         \\
61 & 13 30 33.91 & +47 13 32.3 & 7.2  &26.39 &481 &         \\
62 & 13 30 11.08 & +47 17 10.7 & 6.3  &26.44 &561 &         \\
63 & 13 29 26.01 & +47 17 11.3 & 7.1  &26.47 &    & T       \\
64 & 13 30 20.00 & +47 13 09.1 & 4.8  &26.94 &607 &         \\
\enddata
\tablenotetext{a}{G: object is probably a background Ly$\alpha$ galaxy (see \S
6.1 for details); N: no emission line detected; S: part of the FCJ statistical
sub-sample used to derive the system's distance; T: planetary is west of
NGC~5195 in a tidal tail.}

\end{deluxetable}

\begin{deluxetable}{cccr}
\tablewidth{0pt} \tablenum{2} \tabletypesize{\footnotesize}
\tablecaption{Velocity Errors} \tablehead {
&\multicolumn{2}{c}{Velocity (\kms)} &\colhead{$\Delta v$} \\
\colhead{ID} &\colhead{Setup 1} &\colhead{Setup 2} &\colhead{(\kms)} }
\startdata
 1  &  527 & 539 & $-13$\\
 3  &  661 & 654 & $ +7$\\
 5  &  665 & 664 & $ +1$\\
11  &  674 & 720 & $-46$\\
25  &  431 & 428 & $ +3$\\
27  &  194 & 191 & $ +3$\\
29  &  680 & 698 & $-19$\\
33  &  477 & 479 & $ -2$\\
37  &  316 & 321 & $ -6$\\
51  &  689 & 693 & $ -5$\\
61  &  493 & 469 & $+24$\\
64  &  610 & 601 & $ +9$\\
\enddata
\end{deluxetable}

\begin{figure} \figurenum{1}
\plottwo{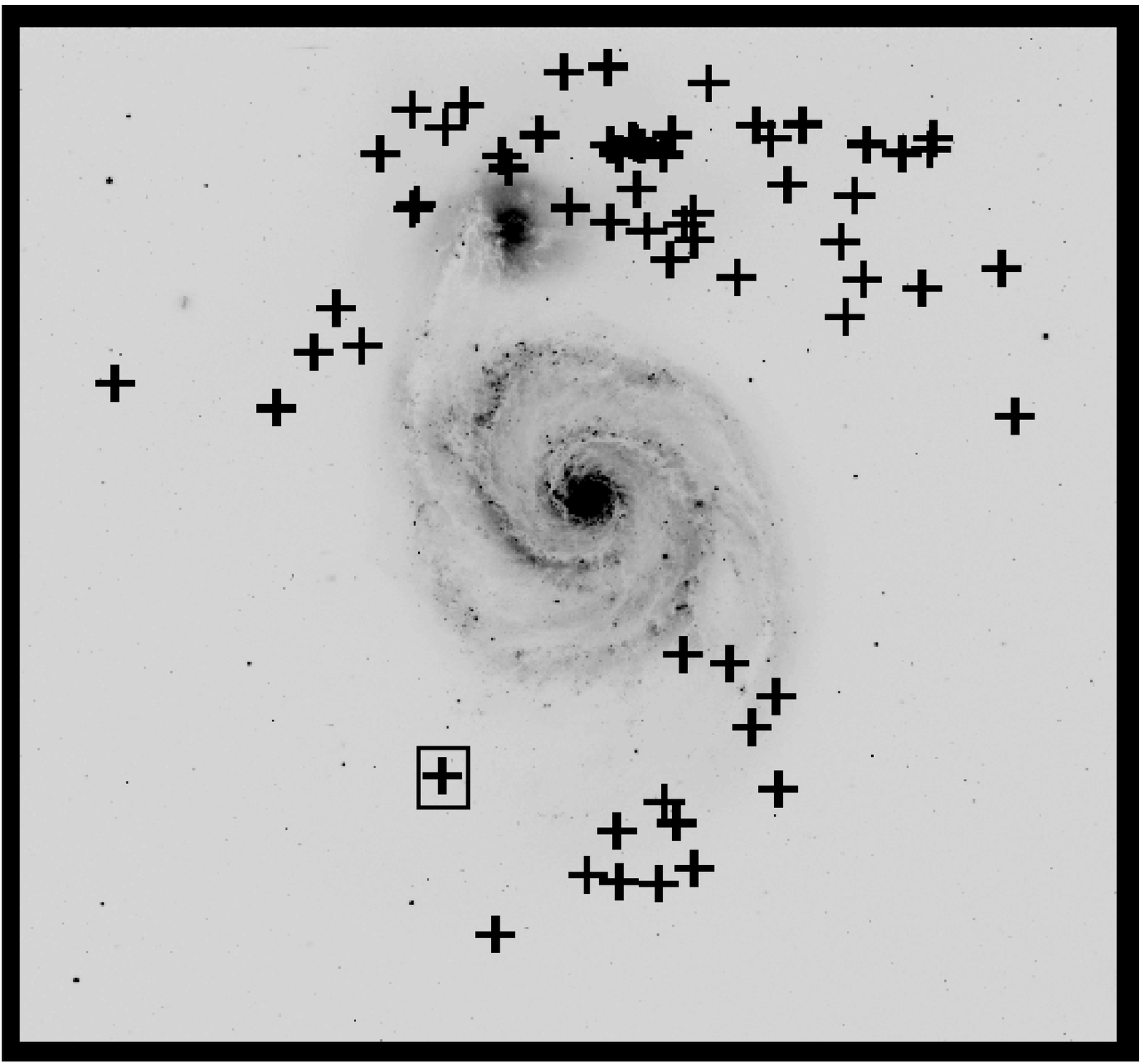}{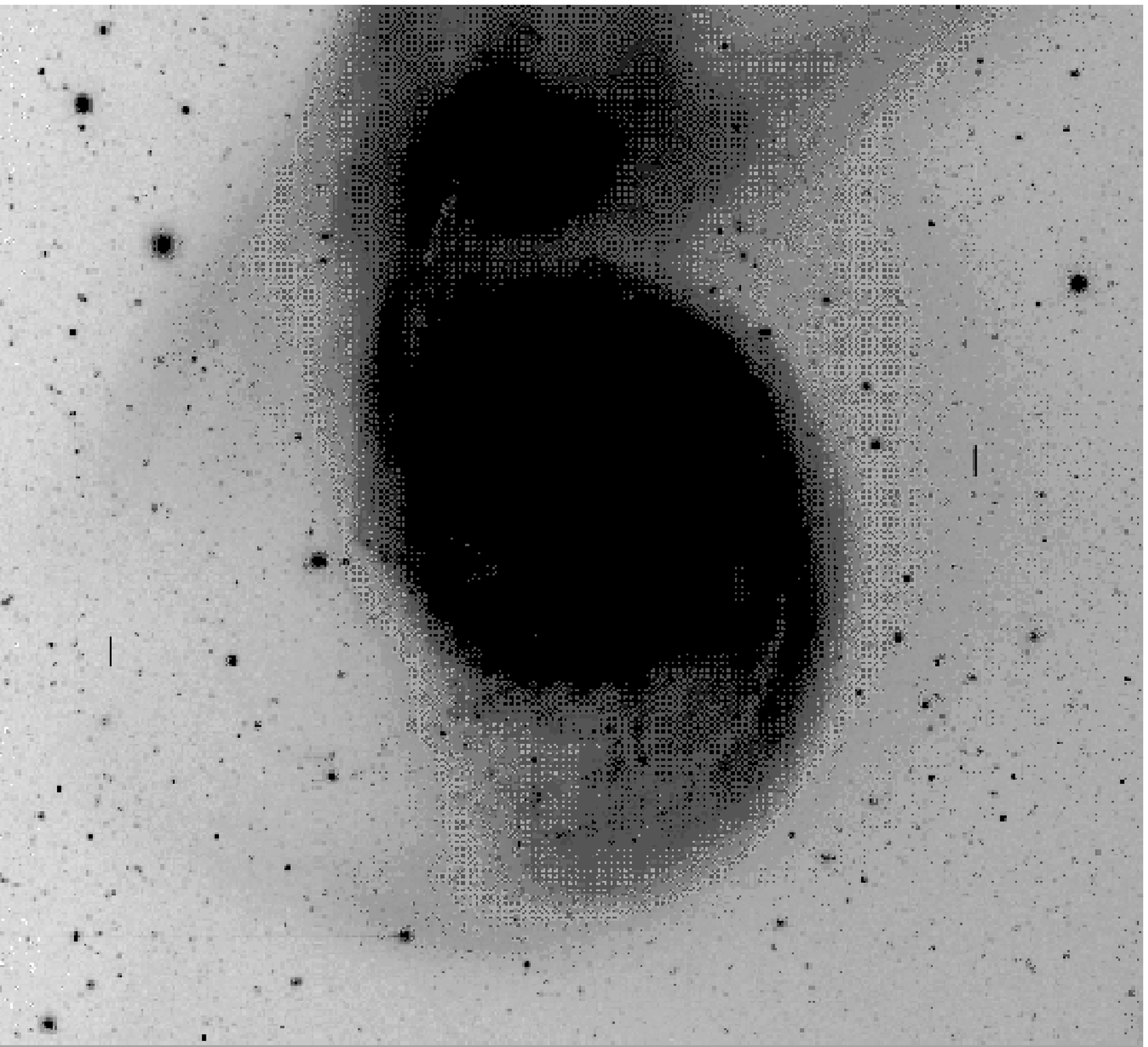} \caption{The left panel displays the
[O~III] $\lambda 5007$ image of M51 taken by \citet{fcj}.  North is up and east
is to the left; the image is $16\arcmin \times 16\arcmin$ in size.  The
positions of 64 planetary nebulae are denoted by crosses; the high-velocity
PN~27 is identified by a box.  On the right is the corresponding off-band
image, binned $5 \times 5$ and logarithmically stretched to bring out low
surface brightness features.  The tidal features described by \citet{zwicky59}
and \citet{burk78} are easily seen.  Note that the planetary nebulae in the
western tail follow a different morphology from the diffuse light.}
\end{figure}

\begin{figure} \figurenum{2(a)}
\plotone{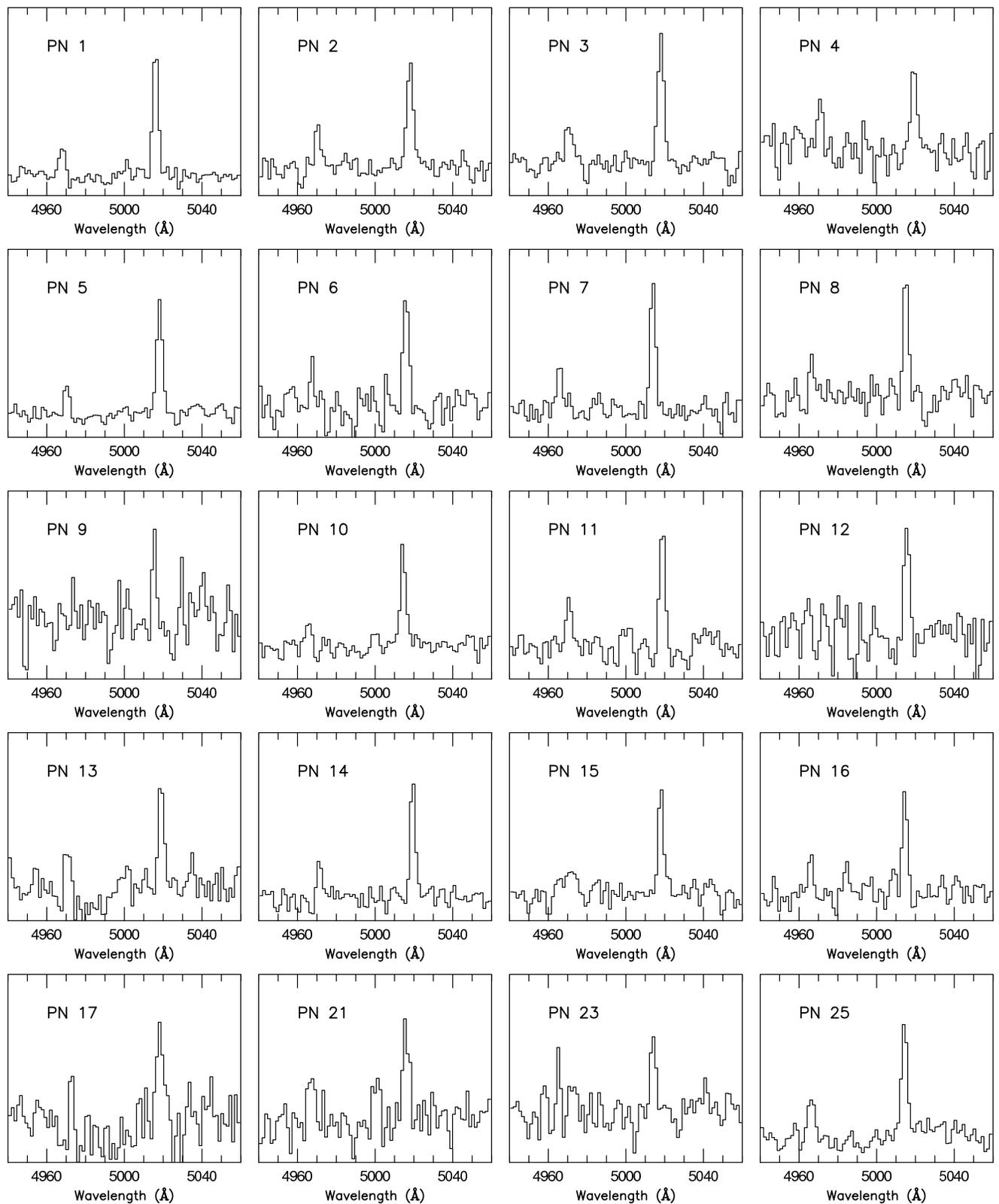} \vskip-1.0truein
\caption{The sky-subtracted spectra of PN candidates 1
through 25 taken with the 3.5~m WIYN telescope.  Only the wavelength region
between 4940 and 5060~\AA\ is shown; the vertical intensity scale is arbitrary.
[O~III] $\lambda 5007$ is clearly visible in each spectrum; [O~III] $\lambda
4959$ is also often visible. \label{fig2a}}
\end{figure}

\begin{figure}\figurenum{2(b)}
\plotone{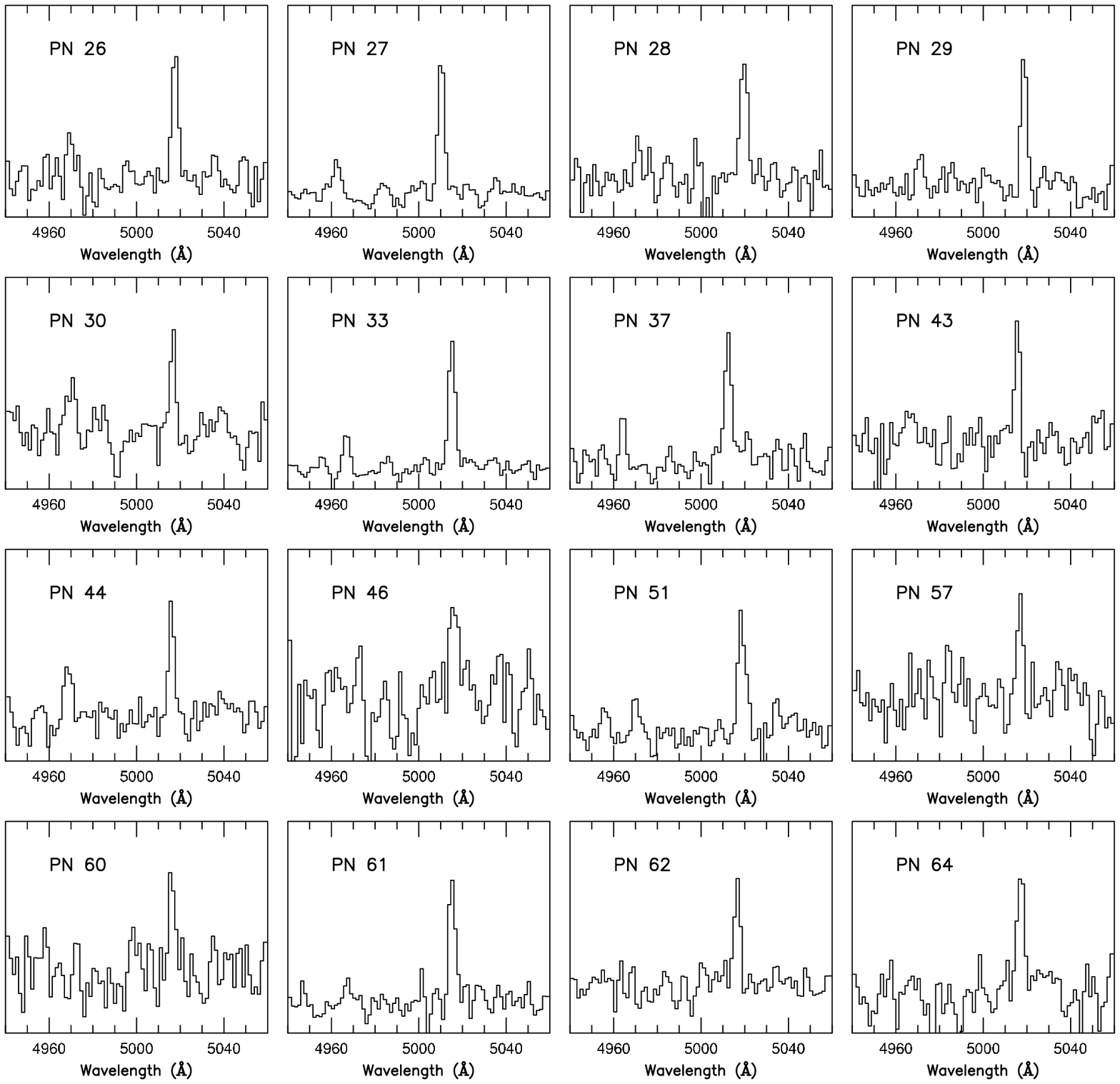} \vskip-1.0truein
\caption{As in Figure 2(a), but for PN candidates 26 through 64. \label{fig2b}}
\end{figure}

\begin{figure}\figurenum{3}
\plotone{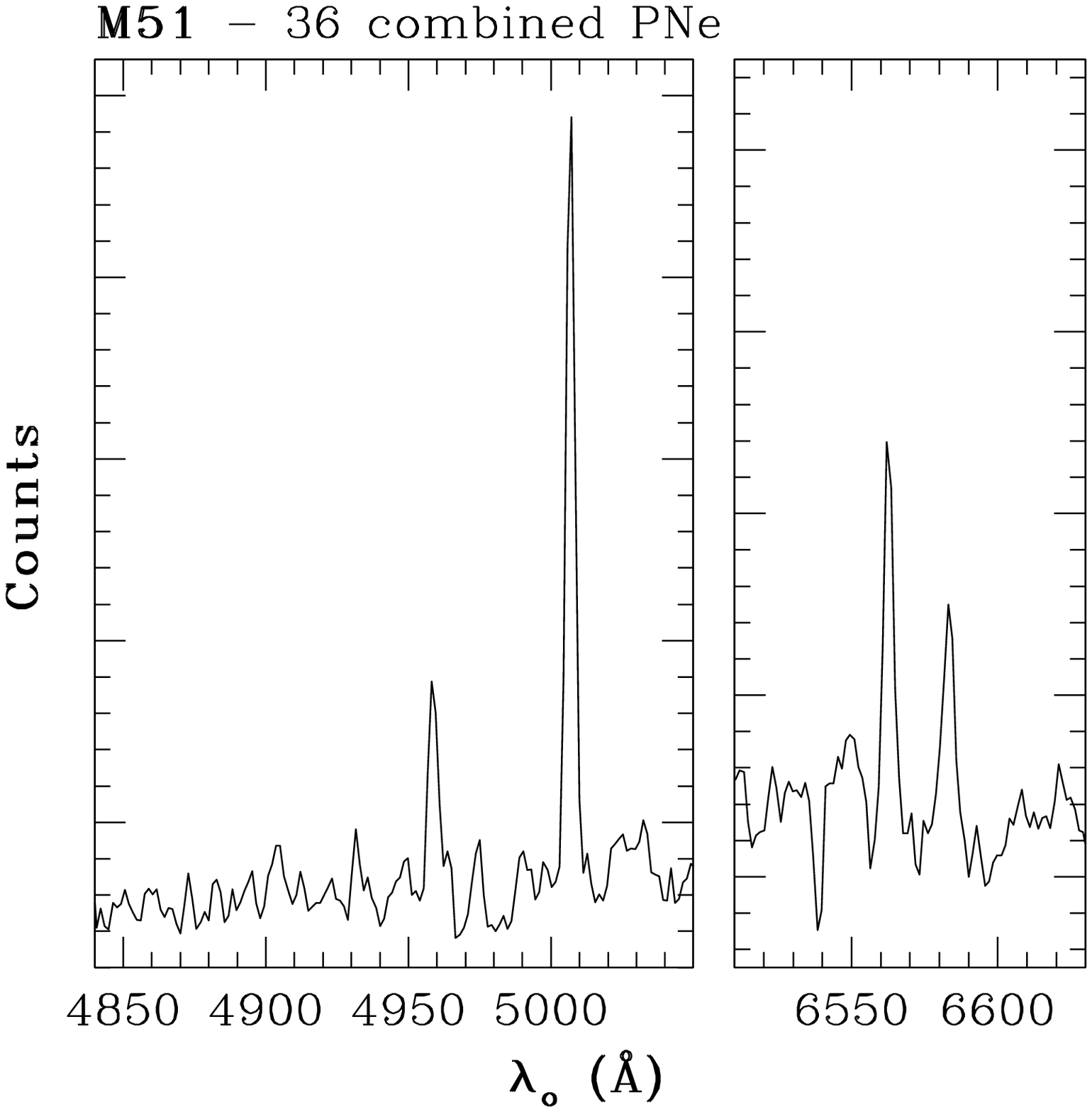} \caption{The co-added sky-subtracted spectra of the 36
candidate M51 PNe, as described in the text.   The left panel shows the region
about [O~III] $\lambda 5007$; the right panel displays the region near
H$\alpha$. As expected for bona-fide planetary nebulae, the [O~III] $\lambda
4959$ line is clearly visible.   The [N~II] $\lambda\lambda 6548,6584$ doublet
is also seen, but the rapid decrease in system thoughput shortward of 5000~\AA\
prevents us from seeing H$\beta$.  The presence of these lines confirms that
the observed objects are genuine planetary nebulae. \label{fig3}}
\end{figure}

\begin{figure}\figurenum{4}
\plotone{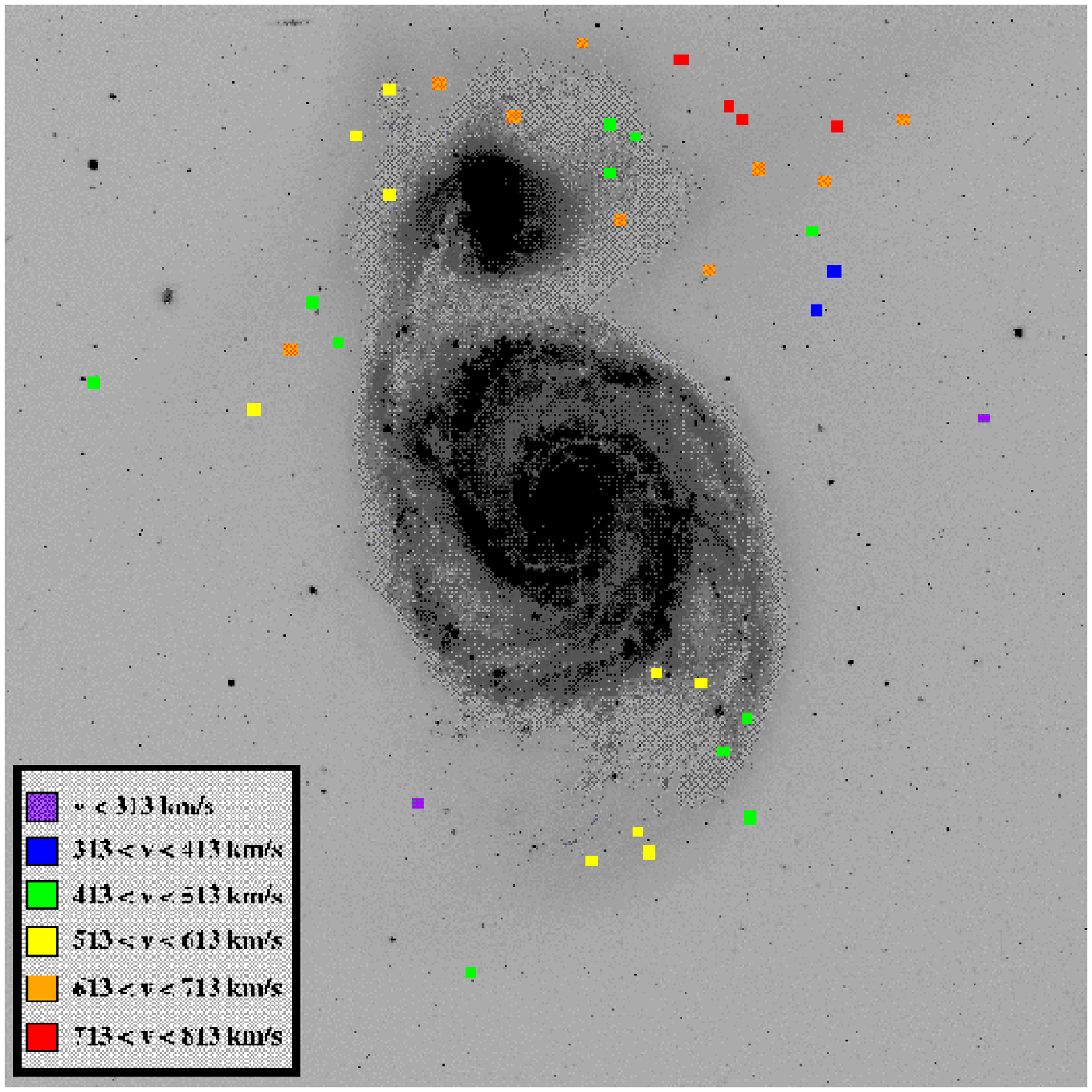} \vskip+0.2truein\caption{The locations of our spectroscopically
confirmed PNe superposed on the FCJ [O~III] $\lambda 5007$ image of M51.  North
is up, and east is to the left; the image is $16\arcmin$ on a side.  The PNe
are color-coded by their heliocentric radial velocity. Note that the tidal tail
structure west of NGC~5195 has multiple velocity components.\label{fig4}}
\end{figure}

\begin{figure}\figurenum{5}
\plotone{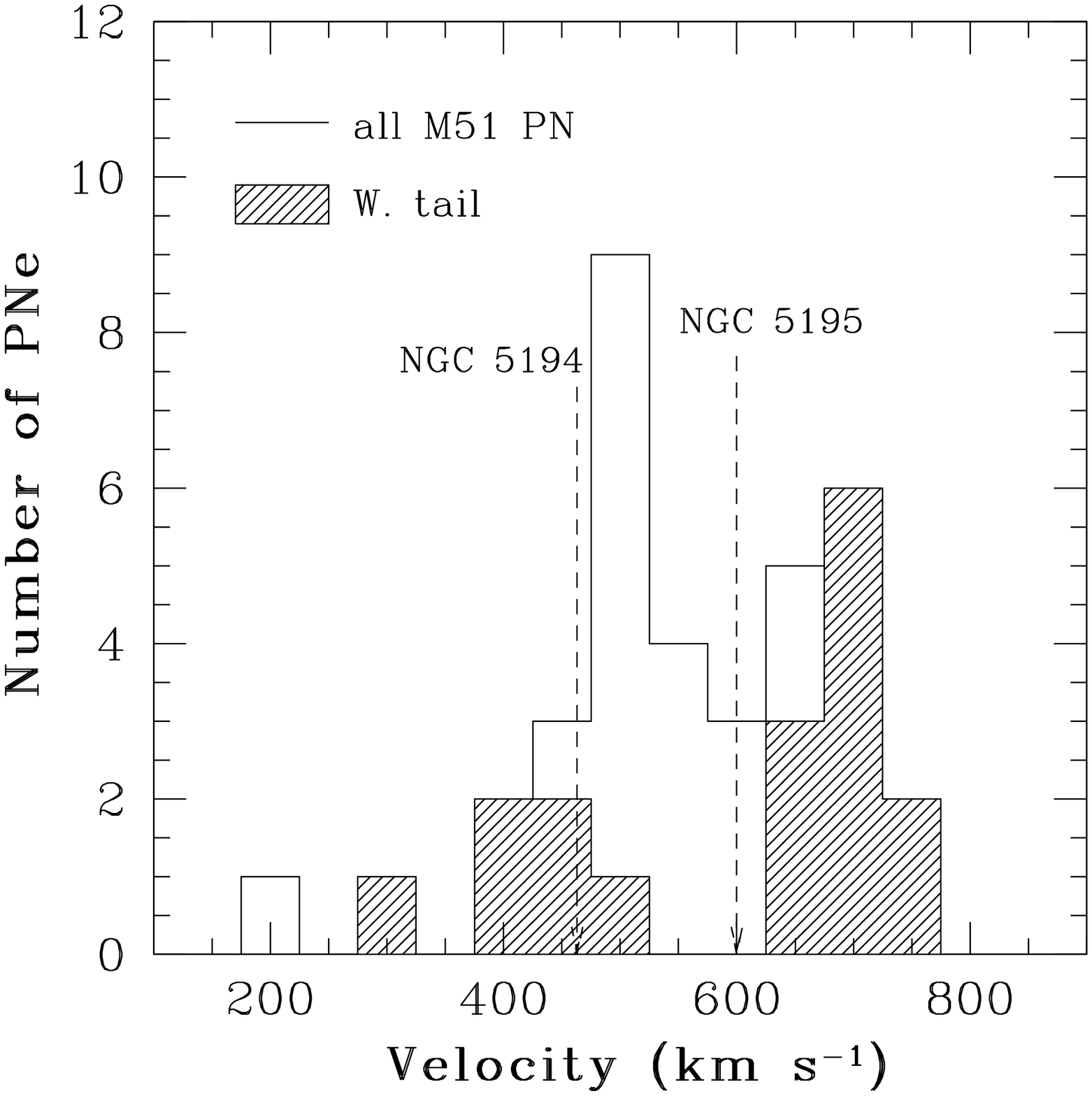} \caption{A histogram of the heliocentric velocities of
PNe in the M51 system.  The hatched region denotes PNe that are located in the
western tail of NGC 5195; see text for details.  The mean heliocentric
velocities for the two galaxies (463~\kms\ for NGC~5194, 600~\kms\ for NGC
5195) are noted.\label{fig5}}
\end{figure}

\begin{figure}\figurenum{6}
\plotone{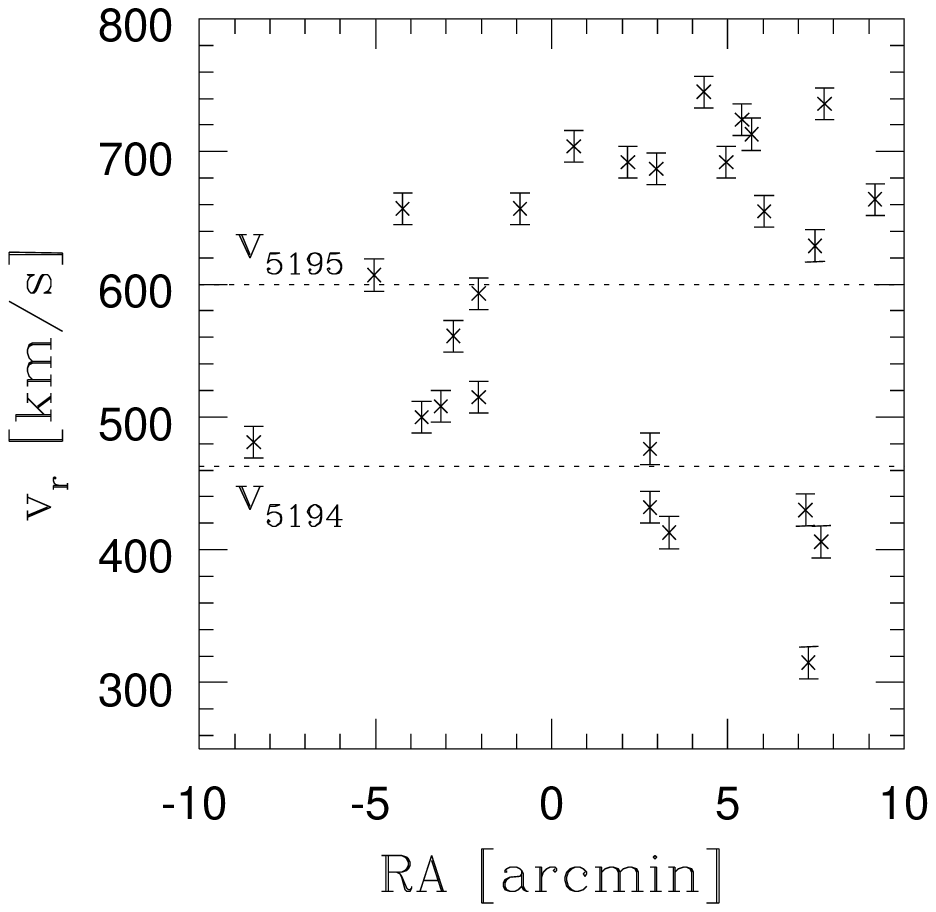} \caption{The velocities of planetary nebulae around
NGC~5195, plotted as a function of right ascension. The spatial zeropoint is
the position of NGC~5195's nucleus.  Multiple dynamically cold components can
be seen in the data.  The bimodal kinematics of the western tail is
obvious.\label{fig6}}
\end{figure}

\begin{figure}\figurenum{7}
\plotone{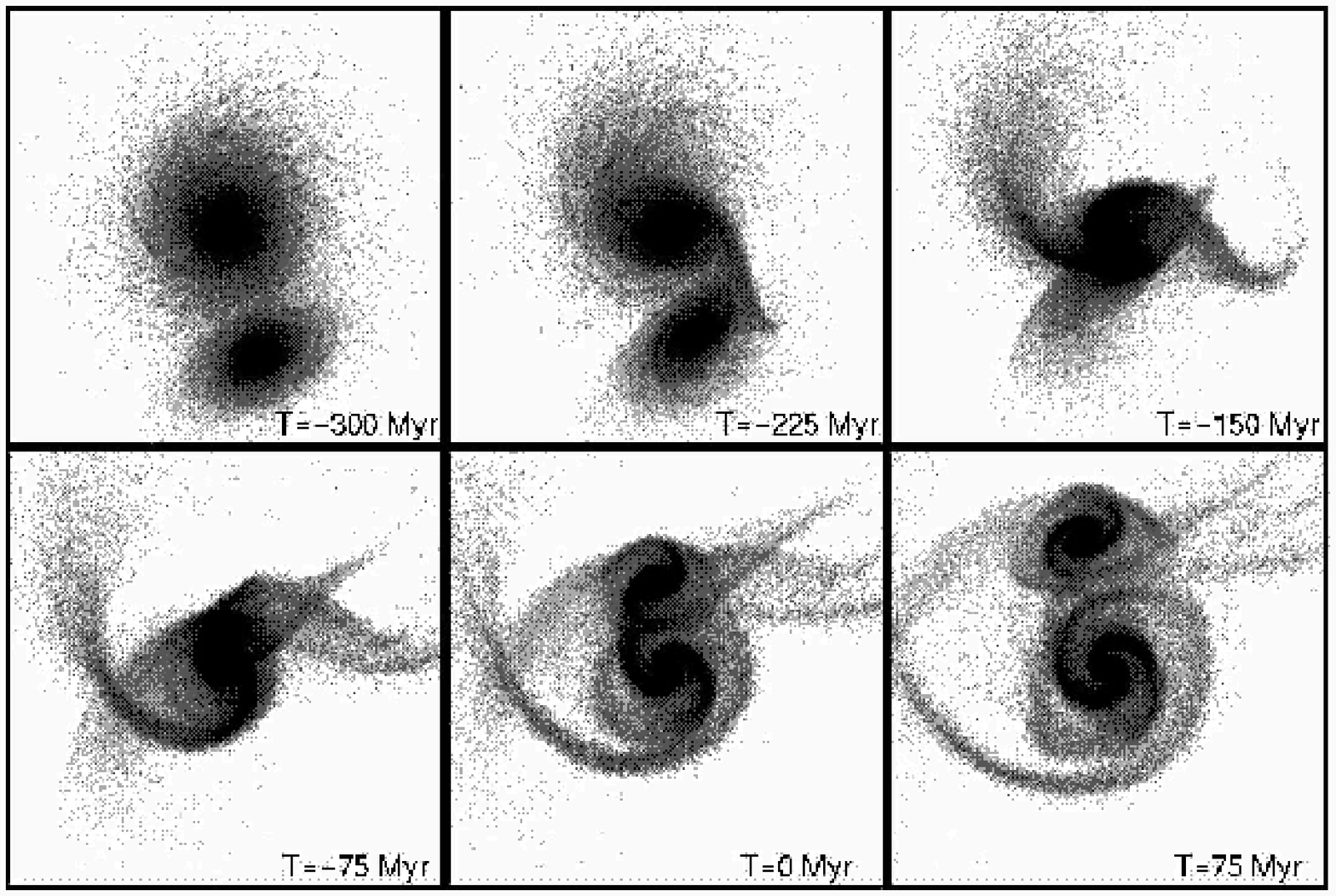} \caption{Time evolution of our new ``single passage''
model, shown projected onto the sky plane. Time is shown in the bottom
right corner of each panel; T=0 corresponds to our best match viewing
time. The sequence starts just before perigalacticon (which occurs at
T=-280 Myr), after which the secondary galaxy, NGC~5195, passes behind
the primary galaxy NGC~5194. At the current time, T=0, the ``western
tail" can be seen to contain tidal material from both galaxies seen in
projection.\label{fig7}}
\end{figure}

\begin{figure}\figurenum{8}
\plotone{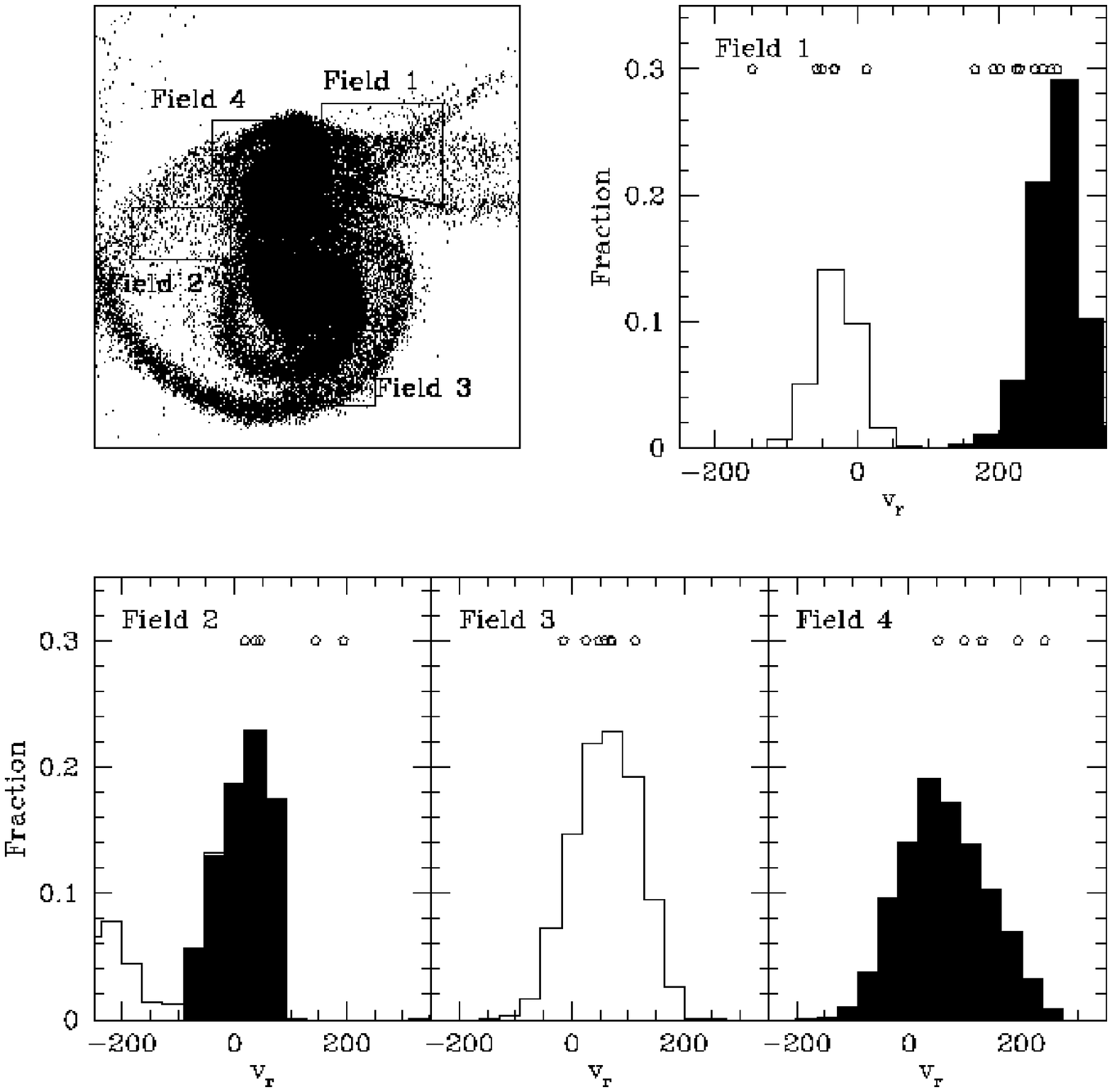} \caption{A comparison between the kinematics of our N-body
model and that observed for PNe in four regions of the M51 system. The velocity
scale is set by the relative velocities of NGC~5194 and 5195
\citep{schweizer77}.  The histograms denote disk N-body particles from NGC~5194
(open histogram) and NGC~5195 (filled histogram); the PNe velocities are
represented by open pentagons.  See the text for a detailed discussion of each
region. \label{fig8}}
\end{figure}

\begin{figure}\figurenum{9}
\plotone{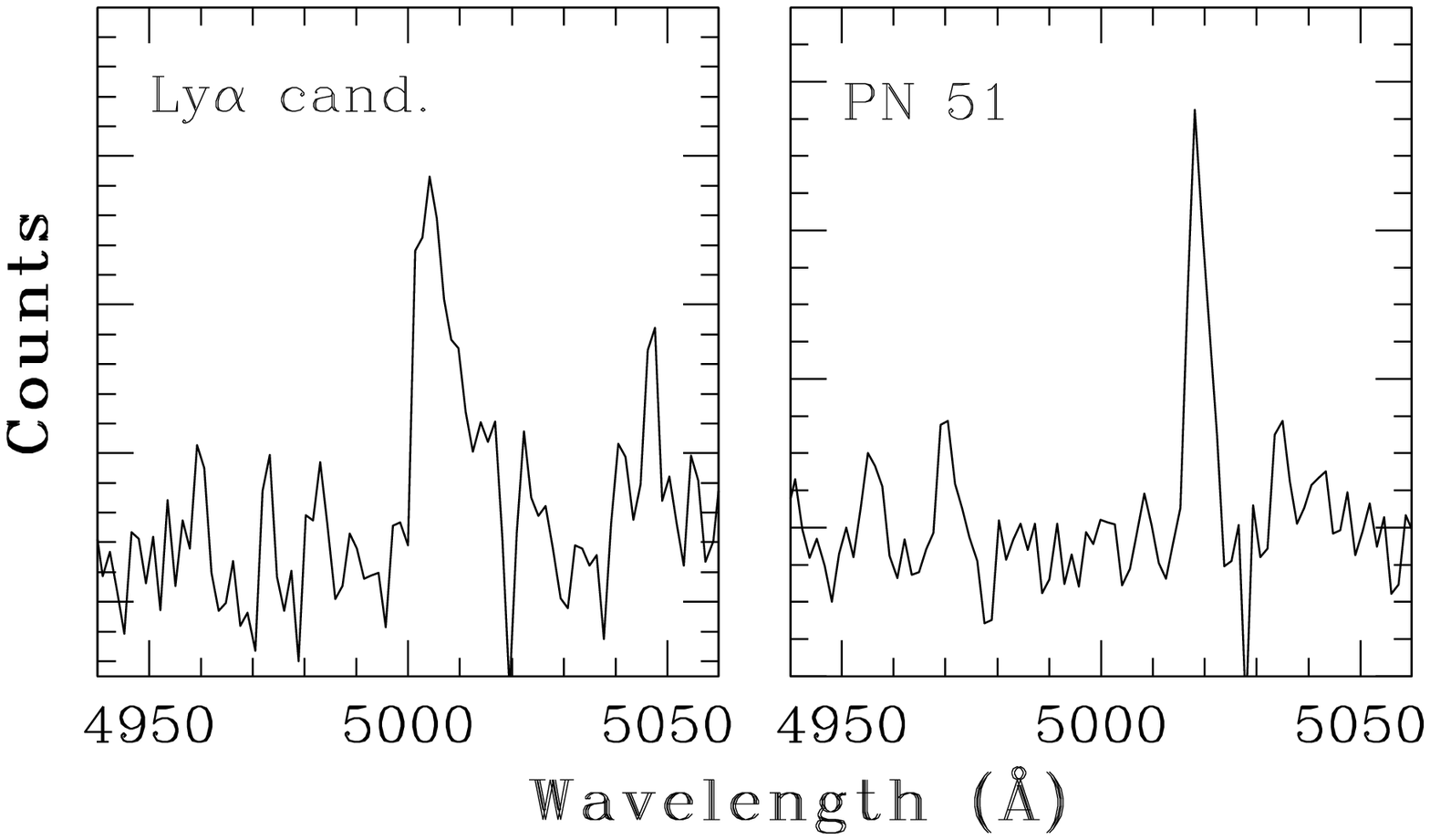} \caption{On the left is our our co-added, sky-subtracted
spectrum of PN candidate 52 between the wavelengths 4940 and 5060~\AA; on the
right is the spectrum of PN~51.  Although the objects are similarly bright in
[O~III] $\lambda 5007$, there are clear differences in their spectral
properties.  We identify PN~52 as a possible Ly$\alpha$ galaxy at
$z=3.12$.\label{fig9}}
\end{figure}

\end{document}